\documentstyle[amssymb,12pt]{article}
\begin{document}

\begin{titlepage}
\begin{center}
{\hbox to\hsize{
\hfill \bf HIP-2001-10/TH}}
{\hbox to\hsize{\hfill  }}

\bigskip
\vspace{3\baselineskip}

{\Large \bf

Extra dimensions and the strong CP problem \\}

\bigskip

\bigskip

{\bf Masud  Chaichian$^{\mathrm{a}}$ and Archil B. Kobakhidze$^{\mathrm{a,b}}$ \\}
\smallskip

{ \small \it
$^{\mathrm{a}}$High Energy Physics Division, Department of Physics, University of Helsinki and \\
Helsinki Institute of Physics, FIN-00014 Helsinki, Finland\\
$^{\mathrm{b}}$Andronikashvili Institute of Physics,GE-380077 Tbilisi, Georgia\\}

\bigskip

\vspace*{.5cm}

{\bf Abstract}\\
\end{center}
\noindent
In higher-dimensional theories such as Brane World models with quasi-localized non-Abelian gauge fields 
the vacuum structure turns out to be trivial. Since the 
gauge theory behaves at large distances as a $4+\delta$-dimensional  and thus 
the topology of the infinity is that of of $S^{3+\delta}$ rather than $S^{3}$, 
the set of gauge mappings are homotopically trivial and the CP-violating 
$\theta$-term vanishes on the brane world-volume. As well there are no contributions to the 
$\theta$-term from the higher-dimensional solitonic configurations. In this way,  
the strong CP problem is absent in the models with quasi-localized gluons.

\bigskip

\bigskip

\end{titlepage}

Despite the fact that the Standard Model (SM) of elementary particles and
their interactions is extremely successful in describing almost all
currently available experimental data, there are some puzzling theoretical
points which force to think that a certain more  fundamental physics stay
behind the SM. One of such puzzles is the strong CP problem. It is reflected 
in the appearance of CP-violating $\theta $-term in the effective Lagrangian
of QCD \cite{1}: 
\begin{equation}
{\cal L}_{\theta }=\frac{\theta }{32\pi ^{2}}Tr\left( G_{\mu \nu }^{a}%
\widetilde{G}^{a\mu \nu }\right) ,  \label{1}
\end{equation}
where $G_{\mu \nu }^{a}$ and $\widetilde{G}^{a\mu \nu }=\frac{1}{2}\epsilon
^{\mu \nu \rho \sigma }G_{\rho \sigma }^{a}$ are gluon field strength and
its dual, respectively. The experimental data on the electric dipole moment
of neutron constraints $\theta $ to be extremely small\footnote{%
More precisely, this bound applies to the effective $\theta $-term defined
as $\overline{\theta }=\theta +agr\det \widehat{M}$, where $\widehat{M}$ is
the quark mass matrix.}, $\theta \lesssim
10^{-10}$, although a priori there are no theoretical reasons
why it should be so tiny.

The existence of the $\theta $-term (\ref{1}) in the effective QCD
Lagrangian is related to the multiple structure of the QCD vacuum. Although (%
\ref{1}) can be expressed as a full derivative it nevertheless does not
vanishe at infinity because of the presence of topologically non-trivial
instanton gauge field configurations \cite{2}. Instantons, being stable, 
finite action solutions to the Euclidean (non-Abelian) gauge field equations
of motion, are interpreted physically as a quantum-mechanical tunnellings
between the different vacuum states $\left| n>\right. $ which are
characterized by a topological index $n\in Z$ \cite{3}. The true vacuum of the 
theory is a superposition of these vacuum states: 
\begin{equation}
\left| \theta >\right. =\sum_{n=-\infty }^{n=\infty }e^{in\theta }\left|
n>\right. .  \label{3}
\end{equation}
The stability of the instantonic configurations is guaranteed by the
non-trivial topology. Indeed, any vacuum configuration of a vector field $%
A_{\mu }$ corresponding to the gauge group $G$ has the form $A_{\mu
}=U^{-1}\partial _{\mu }U$, where $U$ is an element of $G$. Actually, $U$
defines a map from the geometry of a spatial infinity $I$ to a gauge group $G
$: 
\begin{equation}
U:I\rightarrow G.  \label{4}
\end{equation}
In four dimensions $I$ is isomorphic to a 3-sphere $S^{3}$ and if $G$ is a
gauge group which contains an $SU(2)$ as a subgroup, the mapping (\ref{4})
fall into the homotopically non-equivalent classes each characterized by the
topological winding number  $n\in \pi _{3}\left( SU(2)\right) =Z$. The
conservation of the topological charge prevents the instantons from
decaying into the trivial vacuum configuration. This is actually the case
for the colour $SU(3)$ group of QCD. Thus, the strong CP problem emerges due
to the existence of the $\left| n>\right. $-vacua structure in QCD and due
to the non-vanishing amplitude of tunnelling between different $\left| 
n>\right. $-vacua. Consequently, the solution to the strong CP problem can
be achieved either by suppressing the tunnelling or making the $\left|
n>\right. $-vacua unstable. Most of the existing solutions to the strong CP
problem rely on the first possibility. Perhaps the most elegant among them
is the familiar Peccei-Quinn mechanism \cite{4} which actually forbids
tunnellings due to an extra anomalous $U(1)_{PQ}$ setting $\overline{\theta }%
=0$ dynamically. To realize the second possibility one should change the
global geometry of space-time in such a way that mapping (\ref{4}) becomes
homotopically trivial. In this letter we discuss such a possibility in a
particular model with extra dimensions\footnote{% 
For an earlier proposal within a different model see \cite{6}.}.

One can find plenty of purely phenomenomenological motivations why the
extra dimensions beyond those four observed so far could actually be present in 
nature. Moreover, a consistent unification of all fundamental forces within 
the string theory seems also to require introduction of extra dimensions. 
Recently, the idea of extra dimensions received a new twist after the
observation that their effects might be accessible to probe in experiments
in the near future, providing the size of extra dimensions are large enough (%
$\lesssim $TeV$^{-1}$) \cite{7,8} or the extra space is warped \cite{9,10}.
The common ingredient of all these theoretical constructions, often called
as a Brane World scenario, is a (3+1)-dimensional hypersurface (3-brane)
embedded in higher dimensional space-time, where the SM fields \cite{7} and
perhaps gravity as well \cite{9} are confined. Thus the crucial theoretical
question is how the fields of various spin are localized on a 3-brane (for
earlier works see, e.g. \cite{11}). There are significant difficulties in
localizing massless higher-spin fields, and in particular gauge fields, on a
3-brane \cite{11}. In this respect the approach proposed in a series of
recent papers \cite{12,13} is very attractive since it can be universally
applied to all kind of fields.

The basic idea behind this approach is rather transparent \cite{12,13}.
Consider some field freely propagating in the higher-dimensional space-time
(bulk) and having the coupling with another field which is localized on the
3-brane. In most general situations one can expect that the radiative 
corrections with localized field running in the loops can induce the
non-trivial terms (including the kinetic one) for the bulk field on the
3-brane world-volume. Then the emerging physical picture for the bulk field
is the following: At small distances measured on the 3-brane world-volume
the induced 4-dimensional kinetic term dominates over the higher-dimensional 
one and the bulk field essentially behaves as a 4-dimensional one. At large
distances, however, the original higher-dimensional kinetic term becomes
dominant and the field behaves as a higher-dimensional. The crossover scale
is actually controlled by the ratio of the parameters in the original
kinetic term and in that of the induced one (that is the ratio of
higher-dimensional and induced gauge couplings in the case of gauge fields,
for example) and should be adjusted in order not to contradict the already known  
experimental data. This quasi-localization mechanism has been successfully
applied to the gravity \cite{12} and gauge interactions \cite{13} and, as we
mentioned above, can be used for other fields (fermions and bosons) as well.
The remarkable thing offered by this mechanism is that the extra space-time
now can be truly infinite, unlike the case of the ordinary Kaluza-Klein or
warped compactification. This fact can be further explored as a crucial
standpoint for the solution of the cosmological constant problem \cite{12}.
Here we would like to point out another feature of the above scenario with
quasi-localized gauge fields which concerns the strong CP problem. The idea
is due to the above-mentioned observation that the existence of the $\theta $%
-term (\ref{1}) is essentially related to the global geometry of space-time.
In particular, since the gauge theory in the scenario of ref. \cite{13}
become ($4+\delta $)-dimensional at large distances and the topology of
spatial infinity is that of $S^{3+\delta }$ rather than $S^{3}$, the set of
gauge mappings (\ref{4}) become homotopically trivial and the $\theta $-term
(\ref{1}) disappears from the 3-brane world-volume.

To be more quantitative let us begin by considering a pure $SU(2)$ gauge
theory which lives in the 5-dimensional bulk. It is defined by the covariant
derivative $D_{M}=\partial _{M}+iA_{M}^{a}\sigma ^{a}/2$ with field
strengths $G_{MN}=G_{MN}^{a}\sigma ^{a}/2=-i\left[ D_{M},D_{N}\right] $,
where the capital letters run over the bulk coordinates, $M(N)=\left(
\mu (\nu )=0,1,2,3;y\right) $, while those of small, $a,b=1,2,3$ are $SU(2)$%
-adjoint indices and $\sigma ^{a}$ are the Pauli matrices. We assume also that
a (3+1)-dimensional $\delta $-like brane is embedded in 5-dimensional bulk 
space-time with a certain matter fields charged under the $SU(2)$ symmetry 
are localized on it. Then the radiative corrections involving these matter
fields induce the (3+1)-dimensional kinetic term for the $SU(2)$ gauge field
on the 3-brane world-volume, so that the total effective Lagrangian can be written 
as: 
\begin{equation}
{\cal L}=-\frac{1}{2g_{5}^{2}}G_{MN}G^{MN}-\frac{\delta (y)}{2g_{4}^{2}}%
G_{\mu \nu }G^{\mu \nu }+\mbox{...,}  \label{5}
\end{equation}
where $g_{5}$ is the 5-dimensional gauge coupling with mass dimension -1/2
and $g_{4}$ is the dimensionless 4-dimensional gauge coupling and $G_{\mu
\nu }(x^{\mu })=G_{MN}(x^{\mu },y=0)\delta _{\mu }^{M}\delta _{\nu }^{N}$.
Following \cite{13}, we define the crossover scale $r_{c}$ as: 
\begin{equation}
r_{c}=\frac{g_{5}^{2}}{2g_{4}^{2}},  \label{6}
\end{equation}
which need not be extremely large, i.e. of the order of the present Hubble
size, as it could be naively expected, but even $r_{c}\sim 10^{15}$cm (the
solar system size) can be compatible with the existing observations due to
the phenomenon of infrared transparency (see \cite{13} for more details).

Let us first consider two limiting cases: $r_{c}\rightarrow 0$ and $%
r_{c}\rightarrow \infty $. When $r_{c}\rightarrow 0$ ($g_{4}\rightarrow
\infty $, $g_{5}=const$), the induced kinetic term in (\ref{5}) is 
completely irrelevant and thus the $SU(2)$ gauge theory behaves as an exactly 
5-dimensional one. Let us split coordinates as $x^{M}=\left( x^{0},x^{\alpha 
}\right) $, where $x^{\alpha }=\left( x^{i},y\right) $ and $i=1,2,3$ runs
over the spatial 3-brane world-volume coordinates. Now, we can use the
well-known fact of formal mathematical equivalence between the static
solitons in $D$ spatial dimensions and the instantons in $D-1$ space and one
imaginary-time dimensions. So, to construct the static solitonic solutions
in 5 dimensions we must just change the imaginary time coordinate $%
\widetilde{x}^{0}=ix^{0}$ in the original instanton configuration of the
Euclidean 4-dimensional theory by the fifth coordinate $y$. Thus, the
soliton in five dimensions can be written in the form: 
\begin{eqnarray}
A_{0} &=&0,  \nonumber \\
A_{\alpha } &=&-\frac{i}{g_{5}}\eta _{a\alpha \beta }\sigma ^{a}\partial
_{\beta }\ln U,  \label{7}
\end{eqnarray}
where 
\begin{equation}
U(x^{\alpha })=\rho ^{2}+\lambda ^{2},  \label{8}
\end{equation}
$\rho ^{2}=\left( x^{i}-\xi ^{i}\right) ^{2}+\left( y-\xi ^{y}\right)^{2}$, $\lambda $ is the size of the
soliton and $\eta _{a\alpha \beta }$ is the 't Hooft symbol: 
\begin{equation}
\eta _{a\alpha \beta }=\delta _{a\alpha }\delta _{y\beta }-\delta _{a\beta
}\delta _{y\alpha }+\epsilon _{a\alpha \beta }  \label{a1}
\end{equation}
The above configuration describes a static topologically stable soliton 
centered at $x^{\alpha }=\xi ^{\alpha }$ with energy $E=8\pi ^{2}/g_{5}^{2}$ and thus
represents the nontrivial homotopy $\pi _{3}\left( SU(2)\right) $ (just like
the instanton in four dimensions). The configuration (\ref{7}),(\ref{8})
satisfy the self-duality equation of the form 
\begin{equation}
\widetilde{G}_{0\alpha \beta }=G_{\alpha \beta },  \label{a2}
\end{equation}
where $\widetilde{G}_{ABC}=\epsilon _{ABCDE}G^{DE}$. The topological charge $%
q$ is determined through the 4-volume integral over the time component of
the conserved current 
\begin{equation}
Q^{A}=\frac{g_{5}^{2}}{16\pi ^{2}}Tr\widetilde{G}^{ABC}G_{BC}  \label{a3}
\end{equation}
which for (\ref{7}),(\ref{8}) is equal to one: 
\begin{equation}
q=\int \prod_{i=1}^{3}dx^{i}dyQ^{0}=\frac{g_{5}^{2}}{8\pi }E=1.  \label{a4}
\end{equation}

In the opposite extreme case, i.e. when $r_{c}\rightarrow \infty $ ($%
g_{5}\rightarrow \infty $, $g_{4}=const$), the first term in (\ref{5})
disappears and theory becomes (3+1)-dimensional. Passing to the Euclidean 
coordinates ($x^{0}\rightarrow ix^{0}$) we come back to the ordinary
instanton configuration located on the 3-brane world-volume at $y=0$ by
changing $y$ to $x^{0}$ in (\ref{7}),(\ref{8}) with $A_{y}^{a}=0$ instead of $  
A_{0}^{a}=0$ and $\mu =0,1,2,3$ instead of $\alpha =1,2,3,y$.

In the case of finite non-zero $r_{c}$ the gauge fields behave as
(3+1)-dimensional at distances $\left| x^{i}\right| \ll r_{c}$ and as
(4+1)-dimensional at large distances $\left| x^{i}\right| \gg r_{c}$. Hence, 
the spatial infinity is a 4-sphere $S^{4}$ rather than $S^{3}$. Because of
this, the (3+1)-dimensional instanton is not stable against dispersion since 
the gauge mappings $S^{4}\rightarrow SU(2)$ is trivial. That is to say, the
boundary behavior of $SU(2)$ gauge bosons can be always continuously
deformed to the global (trivial) vacuum configuration while keeping the
action finite and therefore the instantons decay into this vacuum. Thus all the 
gauge freedom can be removed by gauge fixing and $\left| n>\right. $-vacua
and the $\theta $-term, consequently, do not arise.

The case of the soliton solution (\ref{7}),(\ref{8}) is somewhat different. In
the case of finite $r_{c}$ the boundary conditions on the 3-brane located at $y=0$ 
become relevant. Since the translational invariance along the fifth coordinate 
is broken by the presence of a 3-brane, the Neumann boundary condition,
\begin{equation}
\left. \frac{\partial A_{\mu }}{\partial y}\right| _{y=0}=0,
  \label{b}
\end{equation}
is satisfied by the one-solitonic configuration (\ref{7}),(\ref{8}) only when 
$\xi ^{y}=0$. One can also construct the multiinstanton-like solitonic solution 
with 
\begin{equation}
U=\biggl ( \frac{\lambda ^{2}}{\rho _{+}^{2}}+
\frac{\lambda ^{2}}{\rho _{-}^{2}}+1 \biggr )
\mbox{ \hspace{0.5cm}}\rho _{\pm
}^{2}=\left( x^{i}\right) ^{2}+(y\mp \xi )^{2}  \label{c}
\end{equation}
in (\ref{7}) instead of (\ref{8}) (see also \cite{14}). This solution describes the static 
soliton centered at $(0,0,0,0,\xi )$ and its image at $(0,0,0,0,-\xi )$ and 
also satisfies the boundary condition (\ref{b}). 
Now, if the fermionic matter (quarks) are localized on the 3-brane, the $%
\theta $-term on the brane will have the form:  
\begin{equation}
{\cal L}_{\theta }=\frac{\theta }{2r_{c}}\int dy\delta (y)Q^{y}.  \label{d}
\end{equation}
It is easy to see that the above solitonic configurations 
indeed lead to the vanishing of (\ref{d}) simply because of $Q^{y}=0$. 

Let us stress that the absence of $\theta $-term is a feature of the
particular models with quasi-localized gauge fields of ref. \cite{13}.
Nothing similar happen in more conventional models with compactified
dimensions where the space-time geometry is a direct product of $M^{4}\times
C^{N}$ ($M^{4}$ is a 4-dimensional Minkowski space-time and $C^{N}$ is an $N$%
-dimensional compactified internal space). Here we have actually two
mappings: $S^{3}\rightarrow SU(2)$ and $B\rightarrow SU(2)$, where $S^{3}$
is a boundary of a Euclideainized $M^{4}$ while $B$ is a boundary of $C^{N}$%
. In fact both mappings might be non-trivial. Thus at least instantons (and $%
\theta $-term, consequently) will remain in $M^{4}$ although there might exist 
other stable topological defects depending on the geometry of extra
space-time $C^{N}$ \cite{15}.

To conclude, we showed that the QCD vacuum is indeed trivial in a certain
class of Brane World models with quasi-localized gluons and thus
CP-violating $\theta $-term is absent. However, in solving the strong CP 
problem in this way, we are loosing the instantonic solution to the $U(1)_{A}
$ problem (i.e. the origin of the $\eta ^{^{\prime }}$ mass). On the other
hand, in the naive quark model there is a simple and natural explanation of
the fact why the $\eta ^{^{\prime }}$ is much heavier than the pion \cite{16}:
in the $\eta ^{^{\prime }}$, which is an isosinglet, the quark -antiquark pair 
can annihilate into gluons while such an annihilation is absent for the
isovector $\pi $'s. Moreover, as it was argued by Witten and Veneziano 
\cite{17} long ago the instantonic dynamics underlying $\eta ^{^{\prime }}$ 
mass is in conflict with predictions based on the large $N_{c}$ expansion. They have suggested  
that  the true dynamical origin of the $\eta ^{^{\prime }}$ mass would be
the coupling of the $U(1)_{A}$ axial anomaly to the topological charge
associated with confinement-related vacuum fluctuations rather than
instantons. Alternatively, it seems promising to look for intrinsically
higher-dimensional solution of the $U(1)_{A}$ problem within the models of
quasi-localized gauge fields.
\paragraph*{Acknowledgement.} During the preparation of this paper we 
became aware from G. Dvali that he is also working out a similar proposal 
for the solution of the strong CP problem. We thank him for correspondence. 
This work was supported by the Academy of Finland under the Project No. 163394.

\end{document}